\documentclass[10pt, oneside, onecolumn]{article}

\usepackage[a4paper, margin=0.75in]{geometry}
\usepackage{cite}
\usepackage{dblfloatfix}
\usepackage{setspace}
\usepackage{multirow}
\usepackage{array}
\usepackage{import}
\usepackage{tikz}
\usepackage[bookmarks=false]{hyperref}
\usepackage[hyphenbreaks]{breakurl}
\usepackage{graphicx}
\usepackage{xcolor,colortbl}
\usepackage{tabularx}
\usepackage{flushend}
\usepackage{amsmath}
\usepackage{wrapfig}

\mathchardef\mhyphen="2D

\begin{document}

\newcommand{\thresholdScore}{0.384}
\newcommand{\repliesInWindows}{3455}
\newcommand{\totalUsers}{449}
\newcommand{\totalTweets}{26051}
\newcommand{\averageTweetsUser}{58.02}
\newcommand{\minTweetsUser}{1}
\newcommand{\maxTweetsUser}{832}
\newcommand{\highUserCount}{129}
\newcommand{\totalNonTaggedTweetCount}{16650}
\newcommand{\totalReplies}{4192}
\newcommand{\totalRetweets}{5209}
\newcommand{\highNonTaggedTweetCount}{998}
\newcommand{\highRepliesTweetCount}{177}
\newcommand{\highRetweetsTweetCount}{2408}
\newcommand{\lowRetweetCountPct}{54}
\newcommand{\highNonTaggedTweetCountPct}{11}
\newcommand{\highUserCountPct}{29}
\newcommand{\retweetsScoreMean}{0.384}
\newcommand{\retweetsScoreMedian}{0.287}
\newcommand{\retweetsScoreStd}{0.282}
\newcommand{\repliesScoreMean}{0.212}
\newcommand{\repliesScoreMedian}{0.200}
\newcommand{\repliesScoreStd}{0.092}
\newcommand{\nonTaggedScoreMean}{0.135}
\newcommand{\nonTaggedScoreMedian}{0.102}
\newcommand{\nonTaggedScoreStd}{0.136}
\newcommand{\windowAverageSize}{5.24}
\newcommand{\windowStdSize}{63.87}
\newcommand{\windowMinSize}{0.01}
\newcommand{\usersScoreZero}{111}
\newcommand{\usersMoreThanTenPct}{36}
\newcommand{\usersMoreThanTenPctPct}{8}
\newcommand{\usersZeroHighScoredPct}{71}
\newcommand{\usersAboveLine}{27}
\newcommand{\usersAboveLinePct}{6}

\title{Using Text Similarity to Detect Social Interactions not Captured by Formal Reply Mechanisms\footnote{A final version of this work was published in the 2015 IEEE 11th International Conference on e-Science (e-Science). It can be found in \url{http://dx.doi.org/10.1109/eScience.2015.31}.}}
\author{
Samuel Barbosa\\
Institute of Mathematics and Statistics\\
University of S\~ao Paulo\\
S\~ao Paulo, Brazil\\
Email: sam@ime.usp.br
\and
Roberto M. Cesar-Jr\\
Institute of Mathematics and Statistics\\
University of S\~ao Paulo\\
S\~ao Paulo, Brazil\\
Email: cesar@ime.usp.br
\and
Dan Cosley\\
Department of Information Science\\
Cornell University\\
Ithaca, NY 14853 USA\\
Email: danco@cs.cornell.edu
}

\maketitle

\begin{abstract}
In modeling social interaction online, it is important to understand when people are reacting to each other.  Many systems have explicit indicators of replies, such as threading in discussion forums or replies and retweets in Twitter. However, it is likely these explicit indicators capture only part of people's reactions to each other; thus, computational social science approaches that use them to infer relationships or influence are likely to miss the mark.  This paper explores the problem of detecting non-explicit responses, presenting a new approach that uses tf-idf similarity between a user's own tweets and recent tweets by people they follow. Based on a month's worth of posting data from \totalUsers{} ego networks in Twitter, this method demonstrates that it is likely that at least \highNonTaggedTweetCountPct{}\% of reactions are not captured by the explicit reply and retweet mechanisms.  Further, these uncaptured reactions are not evenly distributed between users: some users, who create replies and retweets without using the official interface mechanisms, are much more responsive to followees than they appear.  This suggests that detecting non-explicit responses is an important consideration in mitigating biases and building more accurate models when using these markers to study social interaction and information diffusion.
\end{abstract}

\maketitle

\newcommand{\joinNameTweet}[2] {
\texttt{\textbf{#1:} #2}
}

\newcommand{\tweetRowFull}[4] {
\stepcounter{rowcount}
\multirow{2}{0.3cm}{ \therowcount }	& 
\multirow{2}{0.7cm}{ $#2$ } 	& 
#3 \\
&& 
\cellcolor{gray!25} #4 \\ \cline{1-3}
}

\newcommand{\tweetTableFull}[2] {
\newcounter{rowcount}
\setcounter{rowcount}{0}
\begin{table*}[!htbp]
	\centering
	\fontsize{6.5pt}{7pt}\selectfont
		\caption{#1}
		\setlength\tabcolsep{0.05cm}
		\begin{tabularx}{\linewidth}{|>{\raggedright\arraybackslash}m{0.3cm}|>{\raggedright\arraybackslash}m{0.7cm}|>{\raggedright\arraybackslash}m{16.5cm}|}
			\cline{1-3}
			\centering\arraybackslash \textbf{\#} & \centering\arraybackslash \textbf{Score} & \centering\arraybackslash \textbf{Retweets} \\ \cline{1-3} 

\tweetRowFull{Retweet}
{1.0}
{\joinNameTweet
{brandonlondon}
{RT @neiltyson: A H R B Q D W E F L M N S X G I J K O P C T V Y U Z -- Gotta love what the alphabet looks like in alphabetical order.}}
{\joinNameTweet
{neiltyson}
{A H R B Q D W E F L M N S X G I J K O P C T V Y U Z -- Gotta love what the alphabet looks like in alphabetical order.}}

\tweetRowFull{Retweet}
{0.834}
{\joinNameTweet
{michael\_palko}
{RT @8\_Semesters: A girlfriend would be great, but I'm already in a pretty committed relationship with alcoholism and bad decisions.}}
{\joinNameTweet
{8\_Semesters}
{A girlfriend would be great, but I'm already in a pretty committed relationship with alcoholism and bad decisions.}}

\tweetRowFull{Retweet}
{0.768}
{\joinNameTweet
{mike\_sprague}
{RT @mshowalter: All the weird horny stuff between Glenn and Maggie on Walking Dead makes me very uncomfortable.}}
{\joinNameTweet
{mshowalter}
{All the weird horny stuff between Glenn and Maggie on Walking Dead makes me very uncomfortable.}}

\tweetRowFull{Retweet}
{0.602}
{\joinNameTweet
{\_\_ShesBrownSKIN}
{RT @\_CarGotThat: Brandywine Came Out With Win, \#TeamBwine}}
{\joinNameTweet
{\_CarGotThat}
{Brandywine Came Out With Win, \#TeamBwine}}

\tweetRowFull{Retweet}
{0.522}
{\joinNameTweet
{Becchappell}
{RT @oliviaaajayne\_: I just love Toy Story, all of them}}
{\joinNameTweet
{oliviaaajayne\_}
{I just love Toy Story, all of them}}

\tweetRowFull{Retweet}
{0.408}
{\joinNameTweet
{\_\_tiki}
{RT @Greektown1921: Welp now you know}}
{\joinNameTweet
{Greektown1921}
{Welp now you know}}

\tweetRowFull{Retweet}
{0.303}
{\joinNameTweet
{terrigolas}
{RT @jam\_bu88: Facebook is down? Oh no, how are cancer and child abuse going to stop without all those likes? :(}}
{\joinNameTweet
{dsilverman}
{RT @CalebGarling: Stop acting like we have ‘rights' on Facebook http://t.co/geE9NjHH}}

\tweetRowFull{Retweet}
{0.248}
{\joinNameTweet
{Becchappell}
{RT @j4kebro: going to be slightly awkward when Jahmene scans the xfactor winner's single in Asda}}
{\joinNameTweet
{bombaytricycle}
{RT @justaholyfooool: Jahmene is gonna be scanning James Arthers CD at ASDA now.. awks}}
\tweetRowFull{Retweet}
{0.132}
{\joinNameTweet
{HollywoodLadyj}
{RT @RealSkipBayliss: RG3 should give Michael Vick a class in scrambling.}}
{\joinNameTweet
{ESPN\_FirstTake}
{RG3 running at 4G!}}

\tweetRowFull{Retweet}
{0.070}
{\joinNameTweet
{davidAmejia}
{RT @Snoopy: It's Monday, Snoopy! http://t.co/asOF9yPA}}
{\joinNameTweet
{AshKetchum151}
{Mondays are like Zubats. Nobody likes Zubats.}}

			\multicolumn{3}{l}{}\\ \cline{1-3}
			\centering\arraybackslash \textbf{\#} & \centering\arraybackslash \textbf{Score} & \centering\arraybackslash \textbf{Replies} \\ \cline{1-3} 

\tweetRowFull{Reply}
{0.768}
{\joinNameTweet
{esterrick}
{RT @StationBistro1: On deck - Next week's soup is White Bean and Smoked Turkey Chili!}}
{\joinNameTweet
{StationBistro1}
{On deck - Next week's soup is White Bean and Smoked Turkey Chili!}}

\tweetRowFull{Reply}
{0.693}
{\joinNameTweet
{Serrae}
{@MollytheGhost @PhantomRat @hollye83 @hockeybychoice @onlymystory @sjopierce @phouse1964 Hate them.}}
{\joinNameTweet
{hollye83}
{@hockeybychoice @onlymystory @PhantomRat @sjopierce @phouse1964 @MollytheGhost @Serrae Hateful. Just hateful.}}

\tweetRowFull{Reply}
{0.573}
{\joinNameTweet
{HectorBesmonte}
{@EmmittWard @jccassiel @hottiemarkie33 @mark\_purdie @Rhino108 @JasonReedyOH420 yeah! thanks emmit! muah! love, Hugs! for you!}}
{\joinNameTweet
{EmmittWard}
{@HectorBesmonte @jccassiel @hottiemarkie33 @mark\_purdie @Rhino108 @JasonReedyOH420 happy birthday hector sending you love and hugs buddy.}}

\tweetRowFull{Reply}
{0.477}
{\joinNameTweet
{VinnyG5}
{@shanmilanowski I wasn't really that drunk that day...I wouldn't get hammered and let you drive with me...but that's a secret so shhhhh}}
{\joinNameTweet
{shanmilanowski}
{RT @VinnyG5: @shanmilanowski lets do that thing were we get drunk and drive around the city while I'm playing my guitar ......again.}}
\tweetRowFull{Reply}
{0.386}
{\joinNameTweet
{RazWorth}
{@SophieRaby yeah! Buzzin}}
{\joinNameTweet
{SophieRaby}
{@RazWorth do you? :(}}

\tweetRowFull{Reply}
{0.283}
{\joinNameTweet
{SarahMcCallumXX}
{@SarahMcCallumXX @mton1996 forgot the x hahaa}}
{\joinNameTweet
{mton1996}
{@SarahMcCallumXX aw hen, I feel for you x}}

\tweetRowFull{Reply}
{0.245}
{\joinNameTweet
{missRaichl}
{@michel\_andness good morning!}}
{\joinNameTweet
{Michel\_andNess}
{Good morning everyone.}}

\tweetRowFull{Reply}
{0.168}
{\joinNameTweet
{Mahalia\_Enares}
{@kiafranklins\_ HAPPY BIRTHDAY!:) xx}}
{\joinNameTweet
{istoleursmartie}
{@Mahalia\_Enares i can be!!}}

\tweetRowFull{Reply}
{0.133}
{\joinNameTweet
{MeganDoesNOLA}
{@KurlyKonfektion Niiiiice...}}
{\joinNameTweet
{KurlyKonfektion}
{@MeganDoesNOLA lmao! I'm gonna put a slice of bacon with/in my drink and see what happens lol}}

\tweetRowFull{Reply}
{0.068}
{\joinNameTweet
{essfardella}
{@Ali\_Diesel\_ There it is.}}
{\joinNameTweet
{Ali\_Diesel\_}
{RT @shkeeber: I am not a slut.    I'm an erection enthusiast.}}

			\multicolumn{3}{l}{}\\ \cline{1-3}
			\centering\arraybackslash \textbf{\#} & \centering\arraybackslash \textbf{Score} & \centering\arraybackslash \textbf{Non-Tagged} \\ \cline{1-3} 

\tweetRowFull{Non-tagged}
{0.920}
{\joinNameTweet
{hypervocal}
{RT @Reuters: FLASH: \#Egypt's Mursi has left presidential palace, two presidency sources say after protesters, police clash outside.}}
{\joinNameTweet
{AntDeRosa}
{RT @Reuters: FLASH: \#Egypt's Mursi has left presidential palace, two presidency sources say after protesters, police clash outside.}}

\tweetRowFull{Non-tagged}
{0.884}
{\joinNameTweet
{hypervocal}
{It's time. RT @whitehouse: hey guys - this is barack.  ready to answer your questions on fiscal cliff \& \#my2k.  Let's get started. -bo}}
{\joinNameTweet
{ethanklapper}
{RT @whitehouse: hey guys - this is barack.  ready to answer your questions on fiscal cliff \& \#my2k.  Let's get started. -bo}}

\tweetRowFull{Non-tagged}
{0.782}
{\joinNameTweet
{Mrjscott}
{A white woman... RT @T\_dot\_Lee: A woman? RT @majic1021: ‘Fresh Prince' Star Alfonso Ribeiro Weds http://t.co/IfT3Zqlr}}
{\joinNameTweet
{T\_dot\_Lee}
{A woman? RT @majic1021: ‘Fresh Prince' Star Alfonso Ribeiro Weds http://t.co/iTrfZfeM}}

\tweetRowFull{Non-tagged}
{0.697}
{\joinNameTweet
{wulan\_kyuufilan}
{RT @WestlifeFansite: hear @nickybyrneoffic on the radio one minute ago!! it was funny :D x}}
{\joinNameTweet
{WestlifeFansite}
{hear @nickybyrneoffic on the radio one minute ago!! it was funny :D x}}

\tweetRowFull{Non-tagged}
{0.579}
{\joinNameTweet
{esterrick}
{RIP Mr Brubeck. Take five. "@annesaurus: Dave Brubeck, jazz icon, dead at 91. http://t.co/ae9UIRmP"}}
{\joinNameTweet
{Supperphilly}
{RT @annesaurus: Dave Brubeck, jazz icon, dead at 91. http://t.co/sOrBOFBR}}

\tweetRowFull{Non-tagged}
{0.443}
{\joinNameTweet
{Zac\_Hartlage14}
{@BadJerry20 OKC traded James Harden}}
{\joinNameTweet
{24\_Jag}
{Why WOULD OKC TRADE JAMES HARDEN????}}

\tweetRowFull{Non-tagged}
{0.359}
{\joinNameTweet
{Serrae}
{(that should have had a link to the Tina and Amy host Golden Globes article. But I'm too lazy to fix it now)}}
{\joinNameTweet
{MichaelAusiello}
{Genius Move: Tina Fey and Amy Poehler to Host 2012 Golden Globe Awards! http://t.co/zdc2hS8F}}

\tweetRowFull{Non-tagged}
{0.275}
{\joinNameTweet
{nicoleoraha}
{@Niallofficial are you excited to come to Australia and meet all your amazing fans like me? ;) xx \#asknialler 11}}
{\joinNameTweet
{ahoynialler}
{@NiallOfficial EXCITED TO COME BACK TO AUSTRALIA, cause we miss you lots xox \#asknialler}}
\tweetRowFull{Non-tagged}
{0.242}
{\joinNameTweet
{Serrae}
{All of @fatherdowling's \#captainhottie pirate puns for \#ouat are perfect. It's the reason we are twitter friends.}}
{\joinNameTweet
{fatherdowling}
{I apologize in advance for inappropriate pirate puns. \#sorryImnotsorry \#OUAT}}

\tweetRowFull{Non-tagged}
{0.139}
{\joinNameTweet
{ESTL63}
{Forever my lady lol}}
{\joinNameTweet
{DaMontesMom\_415}
{@VictoriaLMathis just go get it!! Lol (the devil) aint I? Lol but I would lol}}

		\end{tabularx}
	\label{tab:#2}
\end{table*}
}

\section{Introduction}

Studies on social networks often use actions people take on other people's online content as evidence of social interactions for developing their models. In domains including Usenet \cite{Joyce2006}, Wikipedia \cite{Black2011}, and Facebook \cite{Gilbert2009}, explicit replies are interpreted as evidence of interpersonal interaction and social ties.  These explicit reactions are also used in studies of influence online, such as predicting when an item is likely to be forwarded in Twitter (e.g., \cite{Suh2010,Comarela2012}).

Not all responses, however, are explicitly marked by the system.  For instance, a post that is explicitly threaded as a reply to a particular post in a discussion forum might nevertheless address another post or posts.  In Twitter, the primary focus of this paper, there are buttons for replying to and retweeting another user's tweet---but users might compose a new tweet that references another recently seen without hitting the reply button.  Users might do this for a variety of reasons, from being inspired to write their own post on a topic they see coming up in their feed to using the system in ways not intended by the designer (such as copying and pasting content into a new tweet rather than pressing a retweet button).  

Being able to identify these non-obvious, indirect responses might allow researchers to have a more accurate view of social interaction than explicit mechanisms provide.  This might also improve overall estimates of users' responsiveness to others, for instance,
at the individual level, they might indicate how desirable a user is as a follower: people might wish to have followers who are more likely to redistribute their content.  Aggregating responsiveness of a user's followers at the ego network level could support better estimates of an individual's potential reach or influence \cite{Domingos2001} based on the responsiveness of their followers.  Better responsiveness measures could also improve transmission probabilities in epidemiology-inspired models of diffusion in social networks \cite{Bakshy2012a}. 

This paper assesses the prevalence of non-explicit responses in a dataset drawn from Twitter, using a measure of normalized textual similarity between a user's tweets and recent friends' tweets based on $tf\mhyphen idf$ scores.  Comparing this to the explicit responses provided by the system shows that explicit indicators of response (replies and retweets) in Twitter are in fact associated with high normalized similarity scores.  Choosing conservative score cutoffs for predicting that a tweet is a response and manually inspecting high-scoring tweets that are not marked as responses suggests that explicit indicators miss at least \highNonTaggedTweetCountPct{}\% of reactions. 
Further, this varies between users: some users systematically fail to use formal response mechanisms, meaning that these users are under-represented in studies that rely on explicit indicators of response and under-counted when considering their potential as information spreaders. These results show that the problem of non-explicit responses is an important one with practical implications for understanding interaction and influence online.

Such studies often focus on computational models for predicting retweet behavior.  
For instance, Suh et al. \cite{Suh2010} apply Principal Component Analysis to decompose tweets into a space of characteristics, showing that URLs, hashtags, the number of followers and followees, and the age of the account are correlated with retweet behavior. 
Comarela et al. \cite{Comarela2012} also find that previous responses to the same tweeter, the tweeter's sending rate, and the age of a tweet influence retweeting, proposing two ranking methods for reordering tweets to increase retweeting.  
Petrovic et al. \cite{Petrovic2011} built a \textit{passive-aggressive} classifier for answering that took into consideration social characteristics of the tweets' author as well as tweets' textual features, finding that social features are more informative.  
Peng et al. \cite{Peng2011} used \textit{Conditional Random Fields} to model the probability of how a user retweets a message. 

Other studies look at variations of the problem.  
Artzi et al. \cite{Artzi2012} applied \textit{Multiple Additive Regression-Trees} and \textit{Maximum Entropy Classifiers} to predict both retweets and replies, while Hong et al. \cite{Hong2011} model both the binary question of whether a tweet would be retweeted and the eventual number of retweets a message might accrue.  
Luo et al. \cite{Luo2013} and Wang et al. \cite{Wang2012} approach a similar problem: given a user and their followers, who will retweet a message generated by the user?  Both created classifiers to predict the followers that would retweet a message.  
Liu et al. \cite{Liu2013} studied the social network of questions and answers in \textit{Sina Weibo} looking for characteristics that are associated with a higher number of answers.

These prior works identify a number of useful features that researchers often take into consideration when developing their models.  These include textual features of Tweets, user preferences or characteristics, and features of users' networks including pairwise relationships and graph structure.  Table \ref{tab:characteristics} presents a number of these features and the papers that have used them in response prediction.  This paper's focus on the prevalence of implicit responses complements these works by identifying tweets that, although not marked as a response, are in fact likely to be real responses.  Such tweets would appear as errors or noise to these models; methods for identifying them might improve both these models and our understanding of why these features matter. For instance, account age might turn out to predict retweet behavior mostly because more experienced users are simply more likely to press the retweet button than new users, rather than having a higher innate propensity to retweet.

When trying to identify non-explicit responses, having a model that explains which messages a user is most likely to be interested can be valuable; that is, the problem of understanding these (message, user) relationships is related to the problem of understanding the (message, reaction) relationships.  
The main stream of research related to modeling user interests in Twitter is the feed personalization problem, defined by Berkovsky et al. \cite{Berkovsky2015} as creating mechanisms that promote and optimize exhibition of interesting content (messages or people, for instance) according to each user's particular preferences and context.  
In their survey, they break approaches to feed personalization into three main groups: approaches that consider the pairwise relationship between author and consumer of content, approaches that take into consideration the graph structure of the social network, and approaches that deal with textual information from the users.

As with studies of retweet prediction, feed personalization approaches often use indicators of tie strength as proxies for potential interest.  
Schaal et al. \cite{Schaal2012} measure pairwise user similarity through tf vectors and topic similarity using LDA. 
Goyal et al. \cite{Goyal2010} estimate pairwise influence probability based on the user activity (action log).  There are a wide variety of such features; Gilbert and Karahalios \cite{Gilbert2009} estimate pairwise tie strength based on Facebook data based on over 70 features in categories including intensity, intimacy, duration, reciprocal services, structural, emotional support, and social distance.  

Network structure also plays an important role in feed personalization.  
Uysal et al. \cite{Uysal2011} developed a personalized tweet ranking method based on a retweet metric, useful in reordering feeds or distributing items to users more likely to retweet. 
Paek et al. \cite{Paek2010} asked Facebook users about the perceived importance of items in their timeline, developed classifiers to identify important messages and friends, and studied the predictive power of a number of features including likes, number of comments, presence of links and images, textual information, and shared background information. 
Both the tie strength and network structure approaches rely on explicit interaction as a tool for estimating tie strength; just as with retweet prediction, being able to identify non-explicit responses might improve these models.

Most related to this paper are text-focused approaches.  Text is commonly used in feed personalization, by comparing content similarity of Tweets or users to a user's previous activity.  
Hannon et al. \cite{Hannon2011} developed a system for follower recommendation on Twitter based on $tf\mhyphen idf$ similarity between the users' newsfeeds. 
Burgess et al. \cite{Burgess2013} propose a system to automatically select users when creating lists. The method adopts $tf\mhyphen idf$ to compare content users generated, among other measures and evaluates the performance comparing user-made lists with those generated by the system.  This work informs ours by providing evidence that $tf\mhyphen idf$-based methods are useful in understanding attention and interest.

\begin{table}[htbp]
	\centering
	\caption{Some characteristics from online social networks that are commonly used to model users' behavior.} 
	\tabcolsep=0.11cm
	\singlespacing
	\fontsize{7pt}{8pt}\selectfont
	\begin{tabular}{|>{\raggedright\centering\arraybackslash}m{1.5cm}|m{6.8cm}|}
		\hline
		\textbf{Characteristic} & \centering\arraybackslash \textbf{Description} \\ \hline
		URL 																								& Presence of a link in a tweet. \cite{Artzi2012,Comarela2012,Peng2011,Petrovic2011,Suh2010} \\ \hline
		Number of hashtags 																	& Number of hashtags in a tweet. \cite{Artzi2012,Comarela2012,Peng2011,Petrovic2011} \\ \hline
		Number of mentions 																	& Number of mentions in a tweet. \cite{Artzi2012,Comarela2012,Liu2013,Peng2011,Petrovic2011,Suh2010} \\ \hline
		Number of followers 																& Number of followers of the author. \cite{Artzi2012,Hong2011,Liu2013,Luo2013,Petrovic2011,Suh2010,Wang2012} \\ \hline
		Number of followees 																& Number of followees of the author. \cite{Artzi2012,Hong2011,Luo2013,Petrovic2011,Suh2010,Wang2012} \\ \hline
		Presence in lists 																	& Number of times that an author has been added to lists. \cite{Luo2013,Petrovic2011} \\ \hline
		Verified 																						& If the author has a verified account. \cite{Luo2013,Petrovic2011} \\ \hline
		Ratio of followers over followees												& Ratio $followers/followees$ or its inverse. \cite{Artzi2012,Peng2011} \\ \hline
		N-grams 																						& Presence of possible n-grams in the text. Usually used together with dimensionality reduction methods. \cite{Artzi2012,Petrovic2011} \\ \hline
		Number of Stop Words 																& Number of stop words in the tweet. \cite{Artzi2012} \\ \hline
		Time 																								& Time when the user received the tweet. \cite{Artzi2012,Liu2013} \\ \hline
		Day of week 																				& Day of the week when the user received the tweet. \cite{Artzi2012} \\ \hline
		Time zone 																					& If the author and the receiver of a tweet are in the same time zone. \cite{Luo2013} \\ \hline
		Wait time 																					& Average time a user takes to reply or retweet a message. \cite{Comarela2012,Hong2011} \\ \hline
		Timeline position 													& How many messages on average a user receives between receiving and replying (or retweeting) a tweet. \cite{Comarela2012} \\ \hline
		Tweet age 																					& When the tweet being retweeted was originally created. \cite{Comarela2012,Hong2011} \\ \hline
		Previous interaction 																& If the user has already replied to or retweeted the author in the past. \cite{Comarela2012,Luo2013,Wang2012} \\ \hline
		Author's activity 																	& Absolute number, frequency, or distribution that represents how the author tweets. \cite{Comarela2012,Hong2011,Liu2013,Luo2013,Peng2011,Petrovic2011,Suh2010,Wang2012} \\ \hline
		Followees activity 																	& Absolute number, frequency, or distribution that represents how the followees of the user tweet. \cite{Peng2011} \\ \hline
		Tweet size 																					& Number of characters of the tweet. \cite{Comarela2012,Petrovic2011} \\ \hline
		Author's PageRank 																	& PageRank of the author. \cite{Hong2011,Wang2012} \\ \hline
		Reciprocal links 
																		& If the author and the user follow each other. \cite{Hong2011,Peng2011,Wang2012} \\ \hline
		Reciprocal followers 																& Number of followers that the author and the user share. \cite{Peng2011,Wang2012} \\ \hline
		Reciprocal followees 																& Number of followees that the author and the user share. \cite{Peng2011,Wang2012} \\ \hline
		Reciprocal mentions 																& Number of tweets where the author mentions the user or the user mentions the author. \cite{Peng2011} \\ \hline
		Reciprocal retweets 																& Number of retweets that the author and the user share. \cite{Peng2011} \\ \hline
		Clustering coefficients 															& Clustering coefficients of the network structure. \cite{Hong2011} \\ \hline
		Previously retweeted message 												& If and how many times a message has been retweeted by other users in the past. \cite{Hong2011,Suh2010} \\ \hline
		Author's retweet count 															& How many messages of the author have been previously retweeted. \cite{Hong2011,Peng2011} \\ \hline
		Emoticons					 												& If there is an emoticon in the tweet. \cite{Liu2013} \\ \hline
		Message topic 																			& Topic identification on the message text or topic similarity measures between the author's interests and the message topic. \cite{Liu2013,Luo2013,Peng2011,Wang2012} \\ \hline
		Language 																						& User's profile language. \cite{Petrovic2011,Wang2012} \\ \hline
		Favorite 																						& If the tweet has been marked as a favorite by the author. \cite{Petrovic2011,Suh2010} \\ \hline
		Response 																						& If the message received is an answer to a previous message. \cite{Petrovic2011} \\ \hline
		Account age 																				& Age of the tweet author's account. \cite{Suh2010,Wang2012} \\ \hline
		Trending topics words 															& If the tweet has \textit{trending topics}' terms. \cite{Petrovic2011} \\ \hline
		Reciprocal hashtags 																& Number of hashtags in common that the author and the user shared in the past. \cite{Wang2012} \\ \hline
		Reciprocal URLs																			& Number of URLs in common that the author and the user shared in the past. \cite{Wang2012} \\ \hline
		Number of lists																			& Number of lists that an author created. \cite{Wang2012} \\ \hline
	\end{tabular}
	\label{tab:characteristics}
\end{table}

\section{Reaction Identification}

This section presents the definition of the problem and the method used to attack the identification of non-explicit reactions in Twitter.

When users decide to post a message in Twitter, they might be reacting to some content they saw from one of their followees. The first assumption is that the evidence for these reactions are the textual features in a given tweet by user $u$ and textual features in the set of recent tweets by $u$'s followees. Another assumption is that, if $u$ tweeted in reaction to a followee's message, there should be higher text similarity between that tweet and that message. This work focus on text features, rather than user or network characteristics found in prior work, because they have been shown to be useful while simplifying data collection, computation, and modeling.

This leads to this work's first research question, about whether text similarity has potential for identifying non-explicit responses.
Do explicit responses in fact tend to have high text similarity? If so, what fraction of high-scoring tweets are non-explicit?  And, even when similarity is lower, when might non-explicit responses be present?

The second research question asked is how these non-explicit responses are distributed among users.  Are many users ``invisible'' because, although they appear to be responsive based on scores, their responses are not explicit?  Why are they lost?  Are they naive or low-frequency users who do not know better than to retype or cut and paste or restate? And, is this likely to be important in estimating the overall responsiveness of users?

\subsection{Influence Window}

\begin{figure}[!htbp]
\centering
\fontsize{8pt}{10pt}\selectfont
\ifx\du\undefined
  \newlength{\du}
\fi
\setlength{\du}{15\unitlength}
\begin{tikzpicture}[scale=0.8]
\pgftransformxscale{1.000000}
\pgftransformyscale{-1.000000}
\definecolor{dialinecolor}{rgb}{0.000000, 0.000000, 0.000000}
\pgfsetstrokecolor{dialinecolor}
\definecolor{dialinecolor}{rgb}{1.000000, 1.000000, 1.000000}
\pgfsetfillcolor{dialinecolor}
\pgfsetlinewidth{0.100000\du}
\pgfsetdash{}{0pt}
\pgfsetdash{}{0pt}
\pgfsetmiterjoin
\definecolor{dialinecolor}{rgb}{0.000000, 0.000000, 0.000000}
\pgfsetstrokecolor{dialinecolor}
\draw (2.653020\du,-4.578300\du)--(2.653020\du,3.865635\du)--(13.062939\du,3.865635\du)--(13.062939\du,-4.578300\du)--cycle;
\definecolor{dialinecolor}{rgb}{0.000000, 0.000000, 0.000000}
\pgfsetstrokecolor{dialinecolor}
\node at (7.857980\du,-0.116333\du){};
\pgfsetlinewidth{0.100000\du}
\pgfsetdash{{1.000000\du}{1.000000\du}}{0\du}
\pgfsetdash{{0.300000\du}{0.300000\du}}{0\du}
\pgfsetmiterjoin
\pgfsetbuttcap
\definecolor{dialinecolor}{rgb}{0.000000, 0.000000, 0.000000}
\pgfsetstrokecolor{dialinecolor}
\pgfpathmoveto{\pgfpoint{2.400000\du}{-13.600000\du}}
\pgfpathcurveto{\pgfpoint{2.400000\du}{-16.200000\du}}{\pgfpoint{13.200000\du}{-16.200000\du}}{\pgfpoint{13.200000\du}{-13.600000\du}}
\pgfpathcurveto{\pgfpoint{13.200000\du}{-11.000000\du}}{\pgfpoint{2.400000\du}{-11.000000\du}}{\pgfpoint{2.400000\du}{-13.600000\du}}
\pgfusepath{stroke}
\pgfsetlinewidth{0.100000\du}
\pgfsetdash{}{0pt}
\pgfsetdash{}{0pt}
\pgfsetbuttcap
{
\definecolor{dialinecolor}{rgb}{0.000000, 0.000000, 0.000000}
\pgfsetfillcolor{dialinecolor}
\pgfsetarrowsend{latex}
\definecolor{dialinecolor}{rgb}{0.000000, 0.000000, 0.000000}
\pgfsetstrokecolor{dialinecolor}
\draw (7.054030\du,-9.400460\du)--(4.412435\du,-12.796051\du);
}
\pgfsetlinewidth{0.100000\du}
\pgfsetdash{}{0pt}
\pgfsetdash{}{0pt}
\pgfsetbuttcap
{
\definecolor{dialinecolor}{rgb}{0.000000, 0.000000, 0.000000}
\pgfsetfillcolor{dialinecolor}
\pgfsetarrowsend{latex}
\definecolor{dialinecolor}{rgb}{0.000000, 0.000000, 0.000000}
\pgfsetstrokecolor{dialinecolor}
\draw (7.407290\du,-9.467540\du)--(6.959627\du,-12.635518\du);
}
\pgfsetlinewidth{0.100000\du}
\pgfsetdash{}{0pt}
\pgfsetdash{}{0pt}
\pgfsetbuttcap
{
\definecolor{dialinecolor}{rgb}{0.000000, 0.000000, 0.000000}
\pgfsetfillcolor{dialinecolor}
\pgfsetarrowsend{latex}
\definecolor{dialinecolor}{rgb}{0.000000, 0.000000, 0.000000}
\pgfsetstrokecolor{dialinecolor}
\draw (7.760560\du,-9.400460\du)--(11.165045\du,-12.864822\du);
}
\definecolor{dialinecolor}{rgb}{0.000000, 0.000000, 0.000000}
\pgfsetstrokecolor{dialinecolor}
\node[anchor=west] at (8.908270\du,-13.424100\du){. . .};
\pgfsetlinewidth{0.100000\du}
\pgfsetdash{}{0pt}
\pgfsetdash{}{0pt}
\pgfsetmiterjoin
\definecolor{dialinecolor}{rgb}{0.000000, 0.000000, 0.000000}
\pgfsetstrokecolor{dialinecolor}
\pgfpathellipse{\pgfpoint{3.831402\du}{-13.542929\du}}{\pgfpoint{0.923132\du}{0\du}}{\pgfpoint{0\du}{0.881171\du}}
\pgfusepath{stroke}
\definecolor{dialinecolor}{rgb}{0.000000, 0.000000, 0.000000}
\pgfsetstrokecolor{dialinecolor}
\node at (3.831402\du,-13.302929\du){};
\definecolor{dialinecolor}{rgb}{0.000000, 0.000000, 0.000000}
\pgfsetstrokecolor{dialinecolor}
\node[anchor=west] at (3.212350\du,-13.542900\du){$f_1$};
\pgfsetlinewidth{0.100000\du}
\pgfsetdash{}{0pt}
\pgfsetdash{}{0pt}
\pgfsetmiterjoin
\definecolor{dialinecolor}{rgb}{0.000000, 0.000000, 0.000000}
\pgfsetstrokecolor{dialinecolor}
\pgfpathellipse{\pgfpoint{6.831402\du}{-13.542929\du}}{\pgfpoint{0.923132\du}{0\du}}{\pgfpoint{0\du}{0.881171\du}}
\pgfusepath{stroke}
\definecolor{dialinecolor}{rgb}{0.000000, 0.000000, 0.000000}
\pgfsetstrokecolor{dialinecolor}
\node at (6.831402\du,-13.302929\du){};
\definecolor{dialinecolor}{rgb}{0.000000, 0.000000, 0.000000}
\pgfsetstrokecolor{dialinecolor}
\node[anchor=west] at (6.212350\du,-13.542900\du){$f_2$};
\pgfsetlinewidth{0.100000\du}
\pgfsetdash{}{0pt}
\pgfsetdash{}{0pt}
\pgfsetmiterjoin
\definecolor{dialinecolor}{rgb}{0.000000, 0.000000, 0.000000}
\pgfsetstrokecolor{dialinecolor}
\pgfpathellipse{\pgfpoint{11.831432\du}{-13.542929\du}}{\pgfpoint{0.923132\du}{0\du}}{\pgfpoint{0\du}{0.881171\du}}
\pgfusepath{stroke}
\definecolor{dialinecolor}{rgb}{0.000000, 0.000000, 0.000000}
\pgfsetstrokecolor{dialinecolor}
\node at (11.831432\du,-13.302929\du){};
\definecolor{dialinecolor}{rgb}{0.000000, 0.000000, 0.000000}
\pgfsetstrokecolor{dialinecolor}
\node[anchor=west] at (11.137900\du,-13.542900\du){$f_m$};
\pgfsetlinewidth{0.100000\du}
\pgfsetdash{}{0pt}
\pgfsetdash{}{0pt}
\pgfsetmiterjoin
\definecolor{dialinecolor}{rgb}{0.000000, 0.000000, 0.000000}
\pgfsetstrokecolor{dialinecolor}
\pgfpathellipse{\pgfpoint{7.407292\du}{-8.586369\du}}{\pgfpoint{0.923132\du}{0\du}}{\pgfpoint{0\du}{0.881171\du}}
\pgfusepath{stroke}
\definecolor{dialinecolor}{rgb}{0.000000, 0.000000, 0.000000}
\pgfsetstrokecolor{dialinecolor}
\node at (7.407292\du,-8.346369\du){};
\definecolor{dialinecolor}{rgb}{0.000000, 0.000000, 0.000000}
\pgfsetstrokecolor{dialinecolor}
\node[anchor=west] at (6.910820\du,-8.583810\du){$u$};
\pgfsetlinewidth{0.100000\du}
\pgfsetdash{{1.000000\du}{1.000000\du}}{0\du}
\pgfsetdash{{0.300000\du}{0.300000\du}}{0\du}
\pgfsetbuttcap
{
\definecolor{dialinecolor}{rgb}{0.000000, 0.000000, 0.000000}
\pgfsetfillcolor{dialinecolor}
\pgfsetarrowsend{stealth}
\definecolor{dialinecolor}{rgb}{0.000000, 0.000000, 0.000000}
\pgfsetstrokecolor{dialinecolor}
\pgfpathmoveto{\pgfpoint{2.400279\du}{-13.600459\du}}
\pgfpatharc{212}{147}{12.351908\du and 12.351908\du}
\pgfusepath{stroke}
}
\pgfsetlinewidth{0.100000\du}
\pgfsetdash{}{0pt}
\pgfsetdash{}{0pt}
\pgfsetmiterjoin
\definecolor{dialinecolor}{rgb}{0.000000, 0.000000, 0.000000}
\pgfsetstrokecolor{dialinecolor}
\draw (2.645190\du,-6.869750\du)--(2.645190\du,-4.969750\du)--(13.055109\du,-4.969750\du)--(13.055109\du,-6.869750\du)--cycle;
\definecolor{dialinecolor}{rgb}{0.000000, 0.000000, 0.000000}
\pgfsetstrokecolor{dialinecolor}
\node at (7.850150\du,-5.679750\du){};
\pgfsetlinewidth{0.100000\du}
\pgfsetdash{}{0pt}
\pgfsetdash{}{0pt}
\pgfsetbuttcap
{
\definecolor{dialinecolor}{rgb}{0.000000, 0.000000, 0.000000}
\pgfsetfillcolor{dialinecolor}
\definecolor{dialinecolor}{rgb}{0.000000, 0.000000, 0.000000}
\pgfsetstrokecolor{dialinecolor}
\draw (2.653020\du,-2.467320\du)--(13.062900\du,-2.467320\du);
}
\pgfsetlinewidth{0.100000\du}
\pgfsetdash{}{0pt}
\pgfsetdash{}{0pt}
\pgfsetbuttcap
{
\definecolor{dialinecolor}{rgb}{0.000000, 0.000000, 0.000000}
\pgfsetfillcolor{dialinecolor}
\definecolor{dialinecolor}{rgb}{0.000000, 0.000000, 0.000000}
\pgfsetstrokecolor{dialinecolor}
\draw (2.653020\du,-0.356333\du)--(13.062900\du,-0.356333\du);
}
\pgfsetlinewidth{0.100000\du}
\pgfsetdash{}{0pt}
\pgfsetdash{}{0pt}
\pgfsetbuttcap
{
\definecolor{dialinecolor}{rgb}{0.000000, 0.000000, 0.000000}
\pgfsetfillcolor{dialinecolor}
\definecolor{dialinecolor}{rgb}{0.000000, 0.000000, 0.000000}
\pgfsetstrokecolor{dialinecolor}
\draw (2.653020\du,1.754650\du)--(13.062900\du,1.754650\du);
}
\definecolor{dialinecolor}{rgb}{0.000000, 0.000000, 0.000000}
\pgfsetstrokecolor{dialinecolor}
\node[anchor=west] at (3.278500\du,-5.863530\du){User's $u$ username and tweet $t_i$};
\definecolor{dialinecolor}{rgb}{0.000000, 0.000000, 0.000000}
\pgfsetstrokecolor{dialinecolor}
\node[anchor=west] at (7.857980\du,-0.356333\du){};
\definecolor{dialinecolor}{rgb}{0.000000, 0.000000, 0.000000}
\pgfsetstrokecolor{dialinecolor}
\node[anchor=west] at (4.983670\du,-3.436350\du){Followee tweet $ft_1$};
\definecolor{dialinecolor}{rgb}{0.000000, 0.000000, 0.000000}
\pgfsetstrokecolor{dialinecolor}
\node[anchor=west] at (4.980080\du,-1.418130\du){Followee tweet $ft_2$};
\definecolor{dialinecolor}{rgb}{0.000000, 0.000000, 0.000000}
\pgfsetstrokecolor{dialinecolor}
\node[anchor=west] at (4.991310\du,2.830500\du){Followee tweet $ft_n$};
\pgfsetlinewidth{0.100000\du}
\pgfsetdash{}{0pt}
\pgfsetdash{}{0pt}
\pgfsetmiterjoin
\pgfsetbuttcap
{
\definecolor{dialinecolor}{rgb}{0.000000, 0.000000, 0.000000}
\pgfsetfillcolor{dialinecolor}
\definecolor{dialinecolor}{rgb}{0.000000, 0.000000, 0.000000}
\pgfsetstrokecolor{dialinecolor}
\pgfpathmoveto{\pgfpoint{13.235600\du}{-4.584360\du}}
\pgfpathcurveto{\pgfpoint{14.235600\du}{-4.724360\du}}{\pgfpoint{13.355600\du}{-0.334364\du}}{\pgfpoint{13.655600\du}{-0.334364\du}}
\pgfpathcurveto{\pgfpoint{13.955600\du}{-0.334364\du}}{\pgfpoint{13.955600\du}{-0.334364\du}}{\pgfpoint{13.655600\du}{-0.334364\du}}
\pgfpathcurveto{\pgfpoint{13.355600\du}{-0.334364\du}}{\pgfpoint{14.145600\du}{4.005640\du}}{\pgfpoint{13.235600\du}{3.865640\du}}
\pgfusepath{stroke}
}
\definecolor{dialinecolor}{rgb}{0.000000, 0.000000, 0.000000}
\pgfsetstrokecolor{dialinecolor}
\node[anchor=west] at (13.848700\du,-0.247892\du){Window $w_i$};
\pgfsetlinewidth{0.100000\du}
\pgfsetdash{{1.000000\du}{1.000000\du}}{0\du}
\pgfsetdash{{0.300000\du}{0.300000\du}}{0\du}
\pgfsetbuttcap
{
\definecolor{dialinecolor}{rgb}{0.000000, 0.000000, 0.000000}
\pgfsetfillcolor{dialinecolor}
\pgfsetarrowsstart{stealth}
\definecolor{dialinecolor}{rgb}{0.000000, 0.000000, 0.000000}
\pgfsetstrokecolor{dialinecolor}
\pgfpathmoveto{\pgfpoint{10.452600\du}{-6.869748\du}}
\pgfpatharc{357}{262}{1.864634\du and 1.864634\du}
\pgfusepath{stroke}
}
\definecolor{dialinecolor}{rgb}{0.000000, 0.000000, 0.000000}
\pgfsetstrokecolor{dialinecolor}
\node[anchor=west] at (7.275010\du,0.424998\du){.};
\definecolor{dialinecolor}{rgb}{0.000000, 0.000000, 0.000000}
\pgfsetstrokecolor{dialinecolor}
\node[anchor=west] at (7.275010\du,0.724998\du){.};
\definecolor{dialinecolor}{rgb}{0.000000, 0.000000, 0.000000}
\pgfsetstrokecolor{dialinecolor}
\node[anchor=west] at (7.275010\du,1.025000\du){.};
\end{tikzpicture}

\caption{Construction of the window $w_i$ for the tweet $t_i$. The tweets in the window (in this paper, $n=100$) are those most generated by user $u$'s followees most recently before $t_i$.} \label{fig:fig_window_explanation}
\end{figure}
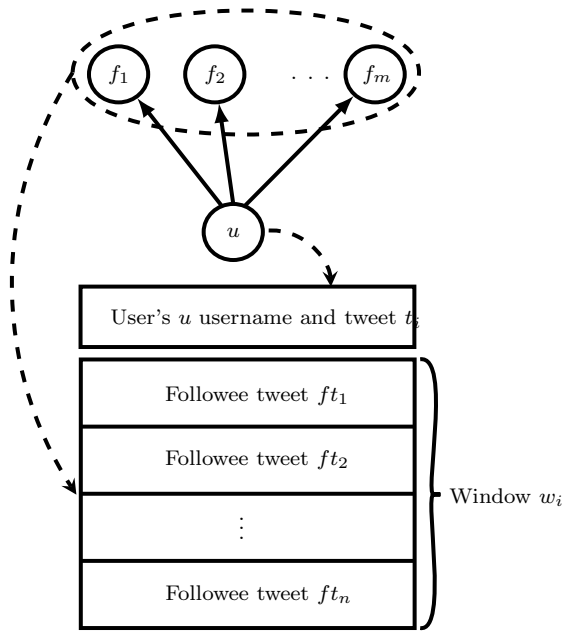

The information Twitter presents to a user is the set of tweets sent by their followees in reverse chronological order. Comarela et al. \cite{Comarela2012} study how far back in the user feed is a tweet when replied or retweeted. They divided the users into four sets of increasing levels of activity and found that over 80\% of replies and 60\% of retweets are responses to one of the 50 most recent tweets in a user's feed. They also present cumulative distributions of these replied and retweeted tweets when varying the position in the feed, and the last 100 tweets in the feed contain more than 80\% of the tweets in these distributions.  
Based on this, a window $w_i$ for the tweet $t_i$ is defined as the last $n=100$ tweets generated by user's $u$ followees $f_i$ immediately before $t_i$, taken in reverse chronological order.
Figure \ref{fig:fig_window_explanation} illustrates the window. 

\subsection{Textual Features}

Each tweet in a user's feed also carries associated meta-data besides the message itself, such as the author's profile and user name, tweet creation time, number of times liked, and number of times retweeted. In this analysis, users are modeled as primarily paying attention to the textual content when considering a response; thus, only textual features the user's feed exposes are considered.  

Tweets are first preprocessed using Python's NLTK package \cite{Bird2009} to be lower case, remove stopwords, and apply Snowball stemming, all common practices when using $tf\mhyphen idf$ scoring.  Hashtags, usernames, and processed words are then extracted using the regular expressions shown in Table \ref{tab:regularExpressions}.  Finally, the tweet author's username is added as a feature since that is also visible in the feed.

\begin{table}[!tb]
	\fontsize{9pt}{10pt}\selectfont
	\centering
		\caption{Regular expressions used to extract features from tweets.}
		\begin{tabular}{|l|c|}
			\hline
			Hashtags & \verb!(?:[\s|^])(#[\w]+)! \\ \hline
			Users & \verb!\B(?:[@＠])([\w]{1,20})! \\ \hline
			Words & \verb!(?:^|[\s][^@＠#\s\w]*)([\w]+)! \\ \hline
		\end{tabular}
	\label{tab:regularExpressions}
\end{table}

\subsection{Message Scoring}

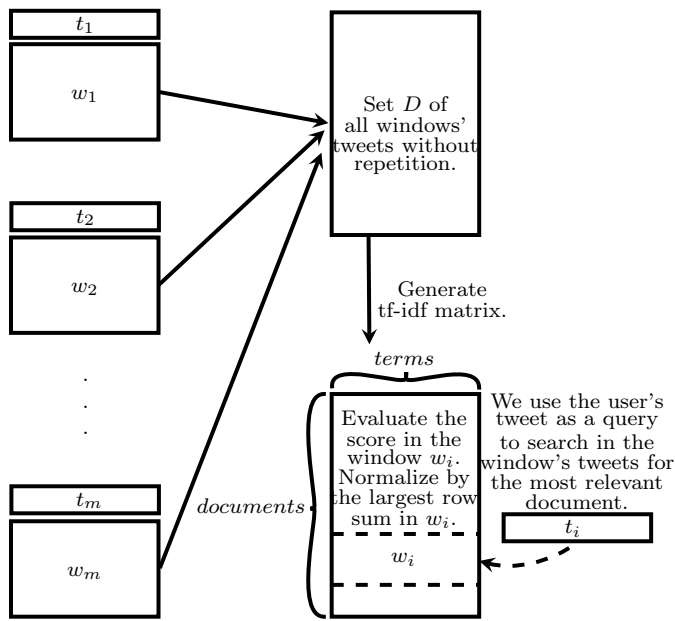
\begin{figure}[!tb]
\centering
\fontsize{8pt}{10pt}\selectfont
\ifx\du\undefined
  \newlength{\du}
\fi
\setlength{\du}{15\unitlength}
\begin{tikzpicture}[scale=0.80]
\pgftransformxscale{1.000000}
\pgftransformyscale{-1.000000}
\definecolor{dialinecolor}{rgb}{0.000000, 0.000000, 0.000000}
\pgfsetstrokecolor{dialinecolor}
\definecolor{dialinecolor}{rgb}{1.000000, 1.000000, 1.000000}
\pgfsetfillcolor{dialinecolor}
\pgfsetlinewidth{0.100000\du}
\pgfsetdash{}{0pt}
\pgfsetdash{}{0pt}
\pgfsetmiterjoin
\definecolor{dialinecolor}{rgb}{0.000000, 0.000000, 0.000000}
\pgfsetstrokecolor{dialinecolor}
\draw (27.982000\du,-15.000000\du)--(27.982000\du,-8.000000\du)--(32.582000\du,-8.000000\du)--(32.582000\du,-15.000000\du)--cycle;
\definecolor{dialinecolor}{rgb}{0.000000, 0.000000, 0.000000}
\pgfsetstrokecolor{dialinecolor}
\node at (30.282000\du,-11.374167\du){};
\pgfsetlinewidth{0.100000\du}
\pgfsetdash{}{0pt}
\pgfsetdash{}{0pt}
\pgfsetmiterjoin
\definecolor{dialinecolor}{rgb}{0.000000, 0.000000, 0.000000}
\pgfsetstrokecolor{dialinecolor}
\draw (17.982000\du,-15.050000\du)--(17.982000\du,-14.200000\du)--(22.582000\du,-14.200000\du)--(22.582000\du,-15.050000\du)--cycle;
\definecolor{dialinecolor}{rgb}{0.000000, 0.000000, 0.000000}
\pgfsetstrokecolor{dialinecolor}
\node at (20.282000\du,-14.499167\du){};
\pgfsetlinewidth{0.100000\du}
\pgfsetdash{}{0pt}
\pgfsetdash{}{0pt}
\pgfsetmiterjoin
\definecolor{dialinecolor}{rgb}{0.000000, 0.000000, 0.000000}
\pgfsetstrokecolor{dialinecolor}
\draw (17.982000\du,-14.000000\du)--(17.982000\du,-11.000000\du)--(22.582000\du,-11.000000\du)--(22.582000\du,-14.000000\du)--cycle;
\definecolor{dialinecolor}{rgb}{0.000000, 0.000000, 0.000000}
\pgfsetstrokecolor{dialinecolor}
\node at (20.282000\du,-12.374167\du){$w_1$};
\pgfsetlinewidth{0.100000\du}
\pgfsetdash{}{0pt}
\pgfsetdash{}{0pt}
\pgfsetmiterjoin
\definecolor{dialinecolor}{rgb}{0.000000, 0.000000, 0.000000}
\pgfsetstrokecolor{dialinecolor}
\draw (17.982000\du,-7.922890\du)--(17.982000\du,-4.922890\du)--(22.582000\du,-4.922890\du)--(22.582000\du,-7.922890\du)--cycle;
\definecolor{dialinecolor}{rgb}{0.000000, 0.000000, 0.000000}
\pgfsetstrokecolor{dialinecolor}
\node at (20.282000\du,-6.297057\du){$w_2$};
\pgfsetlinewidth{0.100000\du}
\pgfsetdash{}{0pt}
\pgfsetdash{}{0pt}
\pgfsetmiterjoin
\definecolor{dialinecolor}{rgb}{0.000000, 0.000000, 0.000000}
\pgfsetstrokecolor{dialinecolor}
\draw (17.982000\du,1.000000\du)--(17.982000\du,4.000000\du)--(22.582000\du,4.000000\du)--(22.582000\du,1.000000\du)--cycle;
\definecolor{dialinecolor}{rgb}{0.000000, 0.000000, 0.000000}
\pgfsetstrokecolor{dialinecolor}
\node at (20.282000\du,2.625833\du){$w_m$};
\definecolor{dialinecolor}{rgb}{0.000000, 0.000000, 0.000000}
\pgfsetstrokecolor{dialinecolor}
\node[anchor=west] at (19.913200\du,-3.460260\du){.};
\definecolor{dialinecolor}{rgb}{0.000000, 0.000000, 0.000000}
\pgfsetstrokecolor{dialinecolor}
\node[anchor=west] at (19.913200\du,-2.613593\du){.};
\definecolor{dialinecolor}{rgb}{0.000000, 0.000000, 0.000000}
\pgfsetstrokecolor{dialinecolor}
\node[anchor=west] at (19.913200\du,-1.766927\du){.};
\pgfsetlinewidth{0.100000\du}
\pgfsetdash{}{0pt}
\pgfsetdash{}{0pt}
\pgfsetbuttcap
{
\definecolor{dialinecolor}{rgb}{0.000000, 0.000000, 0.000000}
\pgfsetfillcolor{dialinecolor}
\pgfsetarrowsend{stealth}
\definecolor{dialinecolor}{rgb}{0.000000, 0.000000, 0.000000}
\pgfsetstrokecolor{dialinecolor}
\draw (22.582000\du,-12.500000\du)--(27.883672\du,-11.518209\du);
}
\pgfsetlinewidth{0.100000\du}
\pgfsetdash{}{0pt}
\pgfsetdash{}{0pt}
\pgfsetbuttcap
{
\definecolor{dialinecolor}{rgb}{0.000000, 0.000000, 0.000000}
\pgfsetfillcolor{dialinecolor}
\pgfsetarrowsend{stealth}
\definecolor{dialinecolor}{rgb}{0.000000, 0.000000, 0.000000}
\pgfsetstrokecolor{dialinecolor}
\draw (22.582000\du,-6.422890\du)--(27.763434\du,-11.294503\du);
}
\pgfsetlinewidth{0.100000\du}
\pgfsetdash{}{0pt}
\pgfsetdash{}{0pt}
\pgfsetbuttcap
{
\definecolor{dialinecolor}{rgb}{0.000000, 0.000000, 0.000000}
\pgfsetfillcolor{dialinecolor}
\pgfsetarrowsend{stealth}
\definecolor{dialinecolor}{rgb}{0.000000, 0.000000, 0.000000}
\pgfsetstrokecolor{dialinecolor}
\draw (22.622657\du,2.475543\du)--(27.662785\du,-10.591456\du);
}
\definecolor{dialinecolor}{rgb}{0.000000, 0.000000, 0.000000}
\pgfsetstrokecolor{dialinecolor}
\node at (30.284900\du,-12.120188\du){Set $D$ of };
\definecolor{dialinecolor}{rgb}{0.000000, 0.000000, 0.000000}
\pgfsetstrokecolor{dialinecolor}
\node at (30.284900\du,-11.485187\du){all windows'};
\definecolor{dialinecolor}{rgb}{0.000000, 0.000000, 0.000000}
\pgfsetstrokecolor{dialinecolor}
\node at (30.284900\du,-10.850187\du){tweets without};
\definecolor{dialinecolor}{rgb}{0.000000, 0.000000, 0.000000}
\pgfsetstrokecolor{dialinecolor}
\node at (30.284900\du,-10.215187\du){repetition. };
\pgfsetlinewidth{0.100000\du}
\pgfsetdash{}{0pt}
\pgfsetdash{}{0pt}
\pgfsetmiterjoin
\definecolor{dialinecolor}{rgb}{0.000000, 0.000000, 0.000000}
\pgfsetstrokecolor{dialinecolor}
\draw (27.982000\du,-3.000000\du)--(27.982000\du,4.000000\du)--(32.582000\du,4.000000\du)--(32.582000\du,-3.000000\du)--cycle;
\definecolor{dialinecolor}{rgb}{0.000000, 0.000000, 0.000000}
\pgfsetstrokecolor{dialinecolor}
\node at (30.282000\du,0.625833\du){};
\pgfsetlinewidth{0.100000\du}
\pgfsetdash{{1.000000\du}{1.000000\du}}{0\du}
\pgfsetdash{{0.300000\du}{0.300000\du}}{0\du}
\pgfsetbuttcap
{
\definecolor{dialinecolor}{rgb}{0.000000, 0.000000, 0.000000}
\pgfsetfillcolor{dialinecolor}
\definecolor{dialinecolor}{rgb}{0.000000, 0.000000, 0.000000}
\pgfsetstrokecolor{dialinecolor}
\draw (27.982000\du,1.400000\du)--(32.582000\du,1.400000\du);
}
\pgfsetlinewidth{0.100000\du}
\pgfsetdash{{0.300000\du}{0.300000\du}}{0\du}
\pgfsetdash{{0.300000\du}{0.300000\du}}{0\du}
\pgfsetbuttcap
{
\definecolor{dialinecolor}{rgb}{0.000000, 0.000000, 0.000000}
\pgfsetfillcolor{dialinecolor}
\definecolor{dialinecolor}{rgb}{0.000000, 0.000000, 0.000000}
\pgfsetstrokecolor{dialinecolor}
\draw (27.982000\du,3.000000\du)--(32.582000\du,3.000000\du);
}
\definecolor{dialinecolor}{rgb}{0.000000, 0.000000, 0.000000}
\pgfsetstrokecolor{dialinecolor}
\node[anchor=west] at (29.533600\du,2.200000\du){$w_i$};
\pgfsetlinewidth{0.100000\du}
\pgfsetdash{}{0pt}
\pgfsetdash{}{0pt}
\pgfsetbuttcap
{
\definecolor{dialinecolor}{rgb}{0.000000, 0.000000, 0.000000}
\pgfsetfillcolor{dialinecolor}
\pgfsetarrowsend{stealth}
\definecolor{dialinecolor}{rgb}{0.000000, 0.000000, 0.000000}
\pgfsetstrokecolor{dialinecolor}
\draw (29.132000\du,-8.000000\du)--(29.158686\du,-4.562557\du);
}
\definecolor{dialinecolor}{rgb}{0.000000, 0.000000, 0.000000}
\pgfsetstrokecolor{dialinecolor}
\node at (31.436600\du,-6.301917\du){Generate};
\definecolor{dialinecolor}{rgb}{0.000000, 0.000000, 0.000000}
\pgfsetstrokecolor{dialinecolor}
\node at (31.436600\du,-5.666917\du){tf-idf matrix.};
\definecolor{dialinecolor}{rgb}{0.000000, 0.000000, 0.000000}
\pgfsetstrokecolor{dialinecolor}
\node at (35.638400\du,-2.788980\du){We use the user's};
\definecolor{dialinecolor}{rgb}{0.000000, 0.000000, 0.000000}
\pgfsetstrokecolor{dialinecolor}
\node at (35.638400\du,-2.153980\du){tweet as a query };
\definecolor{dialinecolor}{rgb}{0.000000, 0.000000, 0.000000}
\pgfsetstrokecolor{dialinecolor}
\node at (35.638400\du,-1.518980\du){to search in the };
\definecolor{dialinecolor}{rgb}{0.000000, 0.000000, 0.000000}
\pgfsetstrokecolor{dialinecolor}
\node at (35.638400\du,-0.883980\du){window's tweets for };
\definecolor{dialinecolor}{rgb}{0.000000, 0.000000, 0.000000}
\pgfsetstrokecolor{dialinecolor}
\node at (35.638400\du,-0.248980\du){the most relevant};
\definecolor{dialinecolor}{rgb}{0.000000, 0.000000, 0.000000}
\pgfsetstrokecolor{dialinecolor}
\node at (35.638400\du,0.386020\du){document.};
\definecolor{dialinecolor}{rgb}{0.000000, 0.000000, 0.000000}
\pgfsetstrokecolor{dialinecolor}
\node at (30.243200\du,-2.242481\du){Evaluate the};
\definecolor{dialinecolor}{rgb}{0.000000, 0.000000, 0.000000}
\pgfsetstrokecolor{dialinecolor}
\node at (30.243200\du,-1.607481\du){score in the};
\definecolor{dialinecolor}{rgb}{0.000000, 0.000000, 0.000000}
\pgfsetstrokecolor{dialinecolor}
\node at (30.243200\du,-0.972481\du){window $w_i$.};
\definecolor{dialinecolor}{rgb}{0.000000, 0.000000, 0.000000}
\pgfsetstrokecolor{dialinecolor}
\node at (30.243200\du,-0.337481\du){Normalize by};
\definecolor{dialinecolor}{rgb}{0.000000, 0.000000, 0.000000}
\pgfsetstrokecolor{dialinecolor}
\node at (30.243200\du,0.297519\du){the largest row};
\definecolor{dialinecolor}{rgb}{0.000000, 0.000000, 0.000000}
\pgfsetstrokecolor{dialinecolor}
\node at (30.243200\du,0.932519\du){sum in $w_i$.};
\pgfsetlinewidth{0.100000\du}
\pgfsetdash{{1.000000\du}{1.000000\du}}{0\du}
\pgfsetdash{{0.300000\du}{0.300000\du}}{0\du}
\pgfsetbuttcap
{
\definecolor{dialinecolor}{rgb}{0.000000, 0.000000, 0.000000}
\pgfsetfillcolor{dialinecolor}
\pgfsetarrowsstart{stealth}
\definecolor{dialinecolor}{rgb}{0.000000, 0.000000, 0.000000}
\pgfsetstrokecolor{dialinecolor}
\pgfpathmoveto{\pgfpoint{32.581740\du}{2.249934\du}}
\pgfpatharc{105}{53}{3.591028\du and 3.591028\du}
\pgfusepath{stroke}
}
\definecolor{dialinecolor}{rgb}{0.000000, 0.000000, 0.000000}
\pgfsetstrokecolor{dialinecolor}
\node[anchor=west] at (19.702600\du,-14.600000\du){$t_1$};
\pgfsetlinewidth{0.100000\du}
\pgfsetdash{}{0pt}
\pgfsetdash{}{0pt}
\pgfsetmiterjoin
\definecolor{dialinecolor}{rgb}{0.000000, 0.000000, 0.000000}
\pgfsetstrokecolor{dialinecolor}
\draw (17.994300\du,-9.000000\du)--(17.994300\du,-8.150000\du)--(22.594300\du,-8.150000\du)--(22.594300\du,-9.000000\du)--cycle;
\definecolor{dialinecolor}{rgb}{0.000000, 0.000000, 0.000000}
\pgfsetstrokecolor{dialinecolor}
\node at (20.294300\du,-8.449167\du){};
\definecolor{dialinecolor}{rgb}{0.000000, 0.000000, 0.000000}
\pgfsetstrokecolor{dialinecolor}
\node[anchor=west] at (19.714900\du,-8.550000\du){$t_2$};
\pgfsetlinewidth{0.100000\du}
\pgfsetdash{}{0pt}
\pgfsetdash{}{0pt}
\pgfsetmiterjoin
\definecolor{dialinecolor}{rgb}{0.000000, 0.000000, 0.000000}
\pgfsetstrokecolor{dialinecolor}
\draw (17.994300\du,-0.100000\du)--(17.994300\du,0.750000\du)--(22.594300\du,0.750000\du)--(22.594300\du,-0.100000\du)--cycle;
\definecolor{dialinecolor}{rgb}{0.000000, 0.000000, 0.000000}
\pgfsetstrokecolor{dialinecolor}
\node at (20.294300\du,0.450833\du){};
\definecolor{dialinecolor}{rgb}{0.000000, 0.000000, 0.000000}
\pgfsetstrokecolor{dialinecolor}
\node[anchor=west] at (19.714900\du,0.350000\du){$t_m$};
\pgfsetlinewidth{0.100000\du}
\pgfsetdash{}{0pt}
\pgfsetdash{}{0pt}
\pgfsetmiterjoin
\definecolor{dialinecolor}{rgb}{0.000000, 0.000000, 0.000000}
\pgfsetstrokecolor{dialinecolor}
\draw (33.343800\du,0.779389\du)--(33.343800\du,1.629389\du)--(37.943800\du,1.629389\du)--(37.943800\du,0.779389\du)--cycle;
\definecolor{dialinecolor}{rgb}{0.000000, 0.000000, 0.000000}
\pgfsetstrokecolor{dialinecolor}
\node at (35.643800\du,1.330222\du){};
\definecolor{dialinecolor}{rgb}{0.000000, 0.000000, 0.000000}
\pgfsetstrokecolor{dialinecolor}
\node[anchor=west] at (34.949800\du,1.229390\du){$t_i$};
\pgfsetlinewidth{0.100000\du}
\pgfsetdash{}{0pt}
\pgfsetdash{}{0pt}
\pgfsetmiterjoin
\pgfsetbuttcap
{
\definecolor{dialinecolor}{rgb}{0.000000, 0.000000, 0.000000}
\pgfsetfillcolor{dialinecolor}
\definecolor{dialinecolor}{rgb}{0.000000, 0.000000, 0.000000}
\pgfsetstrokecolor{dialinecolor}
\pgfpathmoveto{\pgfpoint{27.800000\du}{-3.000000\du}}
\pgfpathcurveto{\pgfpoint{26.720000\du}{-3.120000\du}}{\pgfpoint{27.600000\du}{0.600000\du}}{\pgfpoint{27.420000\du}{0.600000\du}}
\pgfpathcurveto{\pgfpoint{27.240000\du}{0.600000\du}}{\pgfpoint{27.240000\du}{0.600000\du}}{\pgfpoint{27.420000\du}{0.600000\du}}
\pgfpathcurveto{\pgfpoint{27.600000\du}{0.600000\du}}{\pgfpoint{26.720000\du}{4.100000\du}}{\pgfpoint{27.800000\du}{4.000000\du}}
\pgfusepath{stroke}
}
\definecolor{dialinecolor}{rgb}{0.000000, 0.000000, 0.000000}
\pgfsetstrokecolor{dialinecolor}
\node at (25.470271\du,0.507500\du){$documents$};
\pgfsetlinewidth{0.100000\du}
\pgfsetdash{}{0pt}
\pgfsetdash{}{0pt}
\pgfsetmiterjoin
\pgfsetbuttcap
{
\definecolor{dialinecolor}{rgb}{0.000000, 0.000000, 0.000000}
\pgfsetfillcolor{dialinecolor}
\definecolor{dialinecolor}{rgb}{0.000000, 0.000000, 0.000000}
\pgfsetstrokecolor{dialinecolor}
\pgfpathmoveto{\pgfpoint{32.580000\du}{-3.140000\du}}
\pgfpathcurveto{\pgfpoint{32.650000\du}{-4.150000\du}}{\pgfpoint{30.290000\du}{-3.200000\du}}{\pgfpoint{30.300000\du}{-3.500000\du}}
\pgfpathcurveto{\pgfpoint{30.310000\du}{-3.800000\du}}{\pgfpoint{30.310000\du}{-3.800000\du}}{\pgfpoint{30.300000\du}{-3.500000\du}}
\pgfpathcurveto{\pgfpoint{30.290000\du}{-3.200000\du}}{\pgfpoint{27.850000\du}{-4.150000\du}}{\pgfpoint{28.000000\du}{-3.180000\du}}
\pgfusepath{stroke}
}
\definecolor{dialinecolor}{rgb}{0.000000, 0.000000, 0.000000}
\pgfsetstrokecolor{dialinecolor}
\node at (30.244593\du,-4.130328\du){$terms$};
\end{tikzpicture}

\caption{Process to generate each tweet $score$. All the tweets in the windows are used to compose the corpus from which the $tf\mhyphen idf$ matrix for a given user is generated. Each user's tweets are then used as queries to search in their windows for the most relevant followee's tweet.} 
\label{fig:fig_tfidf_explanation}
\end{figure}

The text similarity metric used for this task was the $tf\mhyphen idf$ scoring. It is a proven technique for information retrieval commonly employed in analyzing Twitter data. 
$tf\mhyphen idf$ stands for term frequency and inverse document frequency. This method takes as input a set of documents $D$, where each $document$ is a set of $terms$, and produces a document-by-term matrix of $tf\mhyphen idf$ scores. These functions can be scaled, but usually the $tf$ is not scaled and the $idf$ is logarithmically scaled. For a given $(document,term)$ matrix entry, the $tf$ function is the $term$ occurrence count in the $document$ and the $idf$ function is given by Equation \ref{eq:idf}.

\begin{equation}
idf(D,term) = \log \frac{|D|}{|\left\{d \in D | term \in d\right\}|}
\label{eq:idf}
\end{equation}

Notice here that the $idf$ is a function of the whole set of documents and a particular $term$, while the $tf$ is a function of the document and the $term$.
One high level interpretation of these functions is that $tf$ indicates how important is $term$ for the $document$, while the $idf$ captures how common is the $term$ among the $documents$ and indicates how much information it provides when it occurs in a particular $document$.

The $tf\mhyphen idf$ was calculated using the implementation provided by the Python package scikit-learn \cite{scikit-learn}. It uses a smoothed version of the $idf$ function (even if the $term$ happens in all documents it will not be ignored). The final $tf\mhyphen idf$ document-by-term matrix is given by Equation \ref{eq:tfidf}.

\begin{equation}
tf\mhyphen idf(document,term) = tf*(1-idf)
\label{eq:tfidf}
\end{equation}

The set of documents $D$ is comprised of the tweets in all windows for a user $u$ (each user has its own set $D$, and words in these tweets form a user-specific language model). Each textual feature is one $term$ in our analysis, and the $tf\mhyphen idf$ scores matrix is computed for $D$. 

The tweets generated by $u$ are then used as queries that leverage the matrix.  For each tweet $t_i$, its text features are extracted (removing duplicate $terms$) and the $score$ evaluated for each pair $(t_i,ft_j)$, where $ft_j$ is a followee's tweet in $t_i$'s window $w_i$. The $pairScore$ is given by Equation \ref{eq:score}.

\begin{equation}
pairScore(t_i,ft_j) = \sum_{term \in \left(t_i \cap ft_j\right)}{tf\mhyphen idf(ft_j,term)}
\label{eq:score}
\end{equation}

To be able to compare in a score-independent way between tweets and users, the score for each tweet is normalized based on the maximum value of the $tf\mhyphen idf$ matrix row sum for the tweets in window $w_i$, as given by Equation \ref{eq:normalization}.

\begin{equation}
normalization(w_i) = \max_{t \in w_i}(pairScore(t,t))
\label{eq:normalization}
\end{equation}

This normalization means that the tweet $t_i$ generated by the user will have a final score of $1$ if that tweet reproduces the exact text of the tweet that would yield the maximum score that is present in the window $w_i$. The $score$ for each tweet $t_i$ is then given by Equation \ref{eq:finalScore}.

\begin{equation}
score(t_i) =  \max_{ft_j \in w_i}\frac{pairScore(t_i,ft_j)}{normalization(w_i)}
\label{eq:finalScore}
\end{equation}

The interpretation of the $score(t_i)$ is how likely $t_i$ is to be a response to a friend's tweet $ft = argmax_{ft' \in w_i}(\allowbreak pairScore(t_i,ft'))$.

\section{Twitter Dataset}

This definition of potential response allows ego networks to be collected rather than full network data.  
This is often a more feasible approach when dealing with online social networks, since even friendly APIs normally impose rate limits.  
Ego networks are often useful for studying interaction and influence \cite{Welser2011,Sharma2013}; here, they are appropriate because the method requires only a user's content and his followees' in order to reconstruct the feed windows.

The dataset this paper is based on was collected as part of a project to investigate differences in online behavior between political groups, driven by observations that, in the U.S. 2012 presidential election, Democrats were more active and effective in social media than Republicans.  This paper draws on that dataset, using ego networks on Twitter belonging to users that followed Barack Obama crawled in the first three weeks of December 2012 using V1.0 of the Twitter API.  

The crawler first got all the followers for Obama's account, then filtered out users that did not choose English as their profile language or had no tweets in the last month.  It then randomly selected 547 users and collected up to one month (or the Twitter limit of 3200 historical tweets) of Tweets from each user and all of their followees, creating a set of ego networks.

Because of the 3200 tweet per-user limit, as well as occasional API or network errors, the dataset does not contain a complete record of all followees' tweets.  This could affect estimates of the presence of non-explicit responses; thus, networks where a significant proportion of followees' tweets appeared to be missing were filtered out.  Tweets were considered missing when a followee's activity only partially overlapped with the ego user's\footnote{There is a parallel, opposite problem for users who added followees during the ego user's activity period; windows for tweets before the followee was added will incorrectly contain their tweets, which the user could not have responded to.  We saw no good way to address this and so tolerate the error.}, with the number of missing tweets estimated based on the length of overlap and the rate of that followee's tweets.  Users for whom over 20\% of their followers' tweets were estimated missing were removed from the dataset, leaving \totalUsers{} ego networks\footnote{Other thresholds (5\%, 10\%, 50\%, 80\%, 100\%) were tested.  Lower values lead to similar results, while higher values increased the number of users that lacked data for analysis; 20\% was chosen as a reasonable trade-off between sample size and meaningfulness of results.}.

stribution of the time length for the generated windows.  
Tagged tweets are defined as those indicated by the API as explicit responses, i.e., Replies and Retweets, while the Non-Tagged set is anything not tagged by Twitter\footnote{\label{datasets}Upper-case names refer to the collected sets in this work, while lower-case names refer to messages in general.}. 

\begin{table}[!tb]
	\centering
	\fontsize{9pt}{10pt}\selectfont
		\caption{Descriptive data. Tweet counts are based on ego users. Each tweet may be tagged either as a retweet or a reply. Replies also provide the replied tweet id, allowing us to count how often a tagged reply refers to a tweet in the window.  As with Comarela et al.  \cite{Comarela2012}, over 80\% of tagged replies reference one of the 100 most recent tweets. }
		\begin{tabular}{|l|c|}
			\hline
			Users & \totalUsers{} \\ \hline
			Tweets & \totalTweets{} \\ \hline
			Average Tweets/User & \averageTweetsUser{} \\ \hline
			Min Tweets/User & \minTweetsUser{} \\ \hline
			Max Tweets/User & \maxTweetsUser{} \\ \hline
			Retweets & \totalRetweets{} \\ \hline
			Replies & \totalReplies{} \\ \hline
			Replies in windows & \repliesInWindows{} \\ \hline
			Window avg. size (h) & \windowAverageSize{} \\ \hline
			Windows std. deviation (h) & \windowStdSize{} \\ \hline
			Windows min size (h) & \windowMinSize{} \\ \hline
		\end{tabular}
	\label{tab:databaseInformation}
\end{table}

\begin{figure}[!htbp]
\centering
\includegraphics[scale=1]{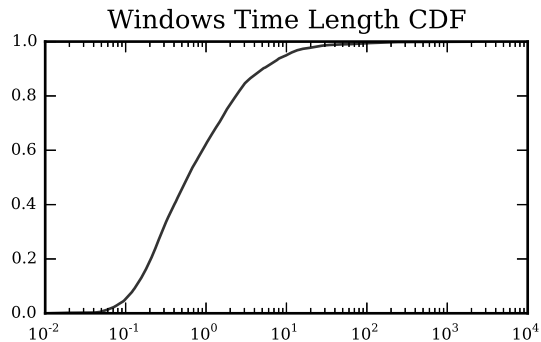}
\caption{Cumulative distribution function for the time length of the windows given in hours. Most windows' lengths are in the interval $[10^{-1}, 10^1]$ hours; about 60\% of windows are 1 hour or less, meaning users receive on average over 100 tweets an hour.}
\label{fig:window_time_size_histogram}
\end{figure}

\subsection{How prevalent are non-explicit responses?}

This section addresses the first research question of whether or not text similarity has potential for identifying untagged responses, starting with whether Tagged reactions indeed tend to have higher scores than Non-Tagged ones. 

\begin{table}[!tb]
	\centering
	\fontsize{9pt}{10pt}\selectfont
		\caption{Sample mean and standard deviation for the normalized similarity score for the Replies, Retweets, and Non-Tagged sets.}
		\begin{tabular}{l|c|c|c|}
			\cline{2-4}
												& Mean					& Median					& Std. \\ \hline 
			\multicolumn{1}{|l|}{Non-Tagged}	& \nonTaggedScoreMean{}	&	\nonTaggedScoreMedian{}	& \nonTaggedScoreStd{} \\ \hline
			\multicolumn{1}{|l|}{Replies}		& \repliesScoreMean{}	&	\repliesScoreMedian{}	& \repliesScoreStd{} \\ \hline
			\multicolumn{1}{|l|}{Retweets}		& \retweetsScoreMean{}	&	\retweetsScoreMedian{}	& \retweetsScoreStd{} \\ \hline
		\end{tabular}
	\label{tab:sampleDistributionsStatistics}
\end{table}

\begin{figure}[!htbp]
\centering
\includegraphics[scale=1]{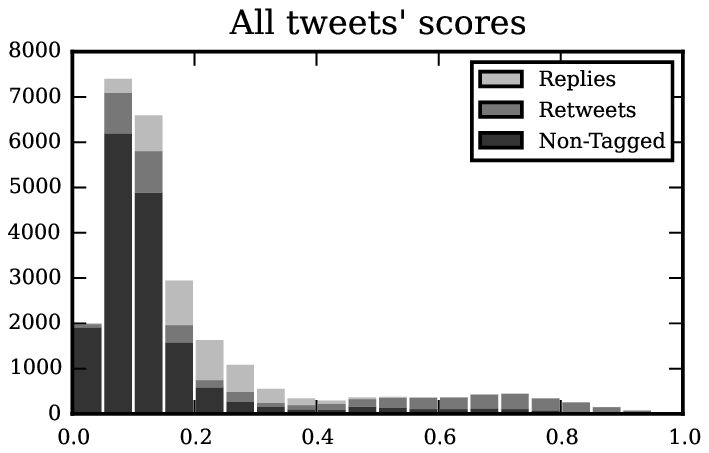}
\includegraphics{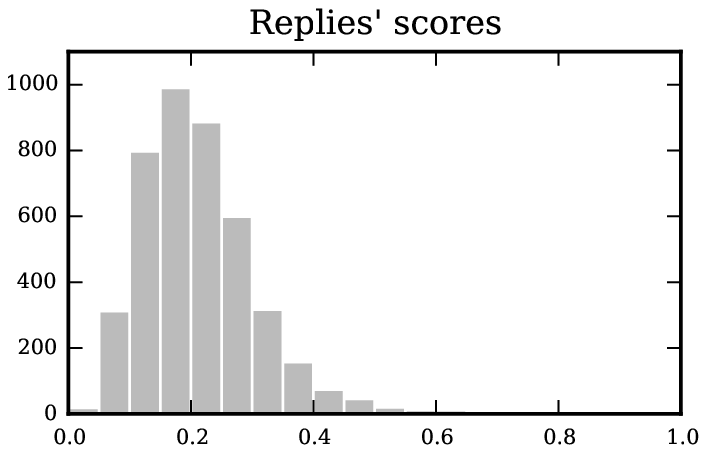}
\includegraphics{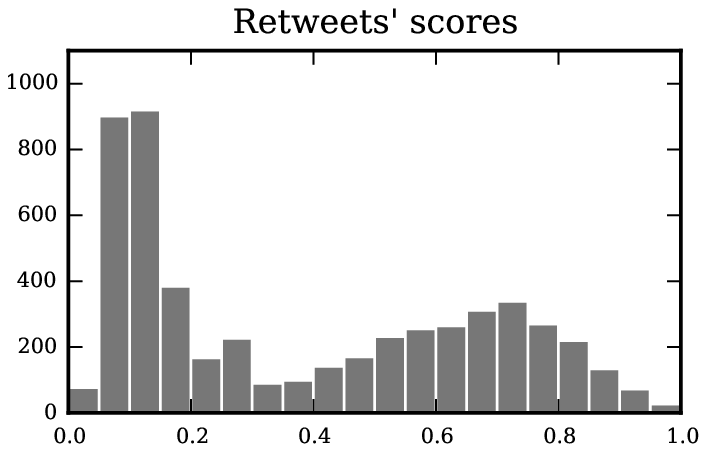}
\includegraphics{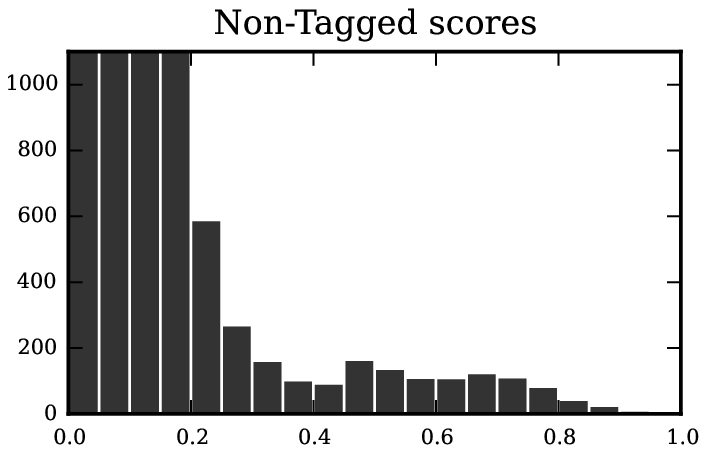}
\caption{Histograms for the normalized similarity scores.  Note that the y-axis for the Non-Tagged subgraph was truncated at 1100 for better visualization of the tail of the distribution and matching other scales.  Retweets have a higher average score than Replies, which in turn are higher than Non-Tagged.  Further, Retweets have a bimodal distribution; high scores are near-duplicates of the tweets they are responding to, but over \lowRetweetCountPct{}\% have a score below \thresholdScore{}, suggesting that people often substantially edit retweets or retweet items not in their feed windows.}
\label{fig:fig_tweets_histograms}
\end{figure}

Mean and median scores are lowest for Non-Tagged and highest for Retweets, as shown in Table \ref{tab:sampleDistributionsStatistics}. This can also be seen in the scores' histogram for each of these sets in Figure \ref{fig:fig_tweets_histograms}. The score behaves as expected when we consider the averages, returning higher values for Replies and Retweets. However, the proximity of the means for the Replies and Non-Tagged and the higher variance of the Non-Tagged makes these two distributions not so well distinguishable based on score alone. The Retweets, on the other hand, present a heavier tail on the distribution. This suggests that the score captures general trends of the Tagged tweets, but is more suitable for Retweets. Considering that the Retweet average is \thresholdScore{} and that it is higher than the Replies mean by more than one standard deviation, \textbf{high scored messages} are defined as messages with $score \geq \thresholdScore{}$.

Although the Non-Tagged set has a lower average, it has a higher variance than replies. This comes from the fact that Non-Tagged tweets have a heavier tail when compared to replies, as seen in Figure \ref{fig:fig_tweets_histograms}. Also, the Non-Tagged high scored tweets are not neglectable when compared with the number of high scored Tagged tweets, as seen in Table \ref{tab:highScoredCounts}: such Non-Tagged tweets would comprise about \highNonTaggedTweetCountPct{}\% of responses, even with a fairly conservative cutoff of \thresholdScore{}. However, high scored messages misses most of the explicit Replies with this cutoff choice.

\begin{table}[!tb]
	\centering
	\fontsize{9pt}{11pt}\selectfont
		\caption{Number of high scored messages and the total of messages for the sets Non-Tagged, Replies and Retweets. The highlighted number of high scored Non-Tagged messages is around \highNonTaggedTweetCountPct{}\% of the highlighted total of Tagged messages.}
		\begin{tabular}{l|>{\centering\arraybackslash}m{1.6cm}|>{\centering\arraybackslash}m{1cm}|>{\centering\arraybackslash}m{1.1cm}|}
			\cline{2-4}
			& Non-Tagged & Replies & Retweets \\ \hline
			\multicolumn{1}{|p{2.2cm}|}{\parbox[top][22pt][c]{2.2cm}{High Scored\\($score \geq \thresholdScore{}$)}} & 
			\cellcolor{gray!25} \highNonTaggedTweetCount{} & \highRepliesTweetCount{} & \highRetweetsTweetCount{} \\ \hline
			\multicolumn{1}{|p{2.2cm}|}{Total} & 
			\totalNonTaggedTweetCount{} & \cellcolor{gray!25} \totalReplies{} & \cellcolor{gray!25} \totalRetweets{} \\ \hline
		\end{tabular}
	\label{tab:highScoredCounts}
\end{table}

Considering the retweet behavior, it would be expected that the normalized similarity score for retweeted messages would be high as long as the original tweet showed up in the windows and the retweet is basically reproducing the message with almost no modifications. 
Surprisingly, this is not what is observed in Figure \ref{fig:fig_tweets_histograms}.  Instead, more than \lowRetweetCountPct{}\% of Retweets have a $score < \thresholdScore{}$.  One possible explanation for this is that people sometimes retweet when they use other parts of the interface, such as other users' profiles or search results, or use social media share buttons attached to tweets on other sites.  Another possibility is that people might frequently edit retweets.

\subsection{Features of Replies, Retweets and Non-Tagged messages}

To help understand the mystery of low-scoring retweets, and more generally to understand what sorts of markers the method is using to identify potential responses, a sample of representative tweets from each category across a range of normalized similarity scores is examined.  
Table \ref{tab:tweetsScores} (see the Appendix) shows both the user's tweet (top in each pair) and the text of the highest-scoring followee's tweet in the window for that tweet (bottom in each pair).

For system tagged Retweets, most of the high scored content has almost the same content as the original message (as expected), as in tweets \#1 and \#2 in the table.  One interesting thing to notice here is that as the tweet length decreases, the normalized similarity score goes down (compare \#6 to \#1). This is related to the fact that the $tf\mhyphen idf$ score is sensitive to the number of matched words between the query and the document.  
Below a threshold of around $0.3$ in this dataset, this effect disappears.  Instead, the text starts to look more like 
two tweets about a common external topic (\#7, \#8, \#9)---despite the fact that the tweet text preserves the ``RT'' retweet marker.  These would be likely candidates for actual retweets that occur outside the window, either farther back in the feed or other parts of the interface than the feed.

When looking at system tagged Replies, high-scoring replies show two main patterns.  In one, they look largely like retweets that were tagged as replies, likely because people pressed the reply button and pasted text from the text they replied to, as in \#11.  In the other, the tweet mentions multiple users who are conducting an ongoing conversation and want all of them to be notified when someone posts something new, as in \#12 and \#13.  It is important to notice that this set of tweets has a maximum score lower than the other sets; scores on the higher end of the distribution could not be found. Also, it appears that @-mentions are the main source of evidence for the normalized similarity scoring even as it goes down, and in fact, replies with low scores still often look like replies despite the low $tf\mhyphen idf$.  This is often (\#16, \#19) but not always (\#18, \#20) indicated by bi-directional @-mentions of the conversational partner.

When looking at Non-Tagged tweets, one of the first things to notice is that high scored tweets usually are retweets that were not captured by the system.  In some cases it is likely users are manually copying the content of the messages and adding retweet markers (\#23, \#24); in others, it is more likely that both users are independently retweeting external content (\#21, \#22).  Users often make small comments together with the original text (\#22, \#23, \#25).  As the normalized similarity score goes down, the messages look less like a retweet, but often still appear to be topically related, sometimes via hashtags (\#28, \#29).  

In general, higher normalized similarity scores seem to capture retweets reasonably well, even though being sensitive to their length, and a particular type of reply that involves conversations.  Non-tagged tweets with high scores are often retweets or quotes with extra comments from the users, although sometimes the retweets may be common retweeting of external content rather than retweets from the window.
Further, even the conservative estimate chosen shows that non-explicit responses are quite common---and it is likely that a number of the of the ``middle scoring'' tweets are actual responses.  Distinguishing those from external influences or underlying interest similarity would be an important next problem in building better models of non-explicit response.  

\subsection{Variations in User Responsiveness}

The previous sections demonstrate that it's likely that \highNonTaggedTweetCountPct{}\% or more of Non-Tagged tweets are responses that are not explicitly captured by the system.  This section addresses the other main research question of how these losses are distributed among different users in the network.   

\begin{figure}[!tb]
\centering
\includegraphics[scale=0.55]{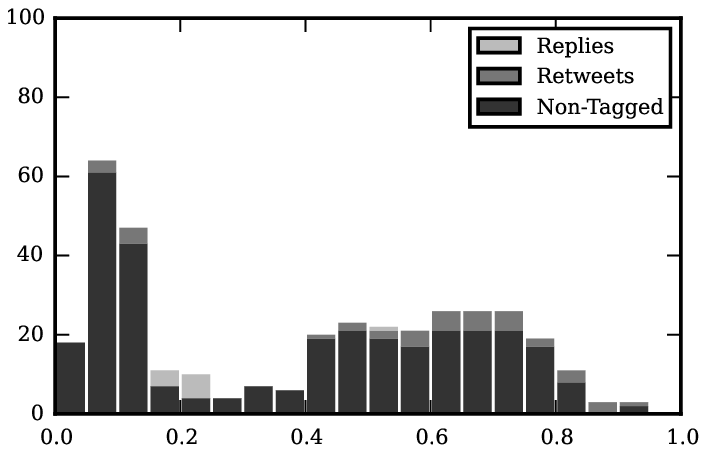}
\includegraphics[scale=0.55]{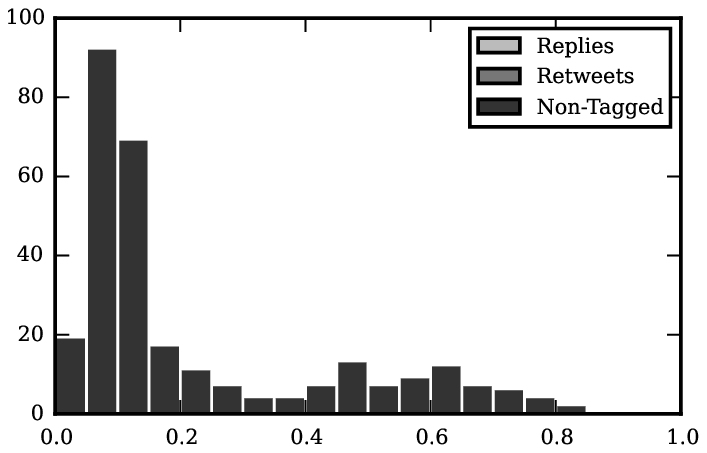}
\includegraphics[scale=0.55]{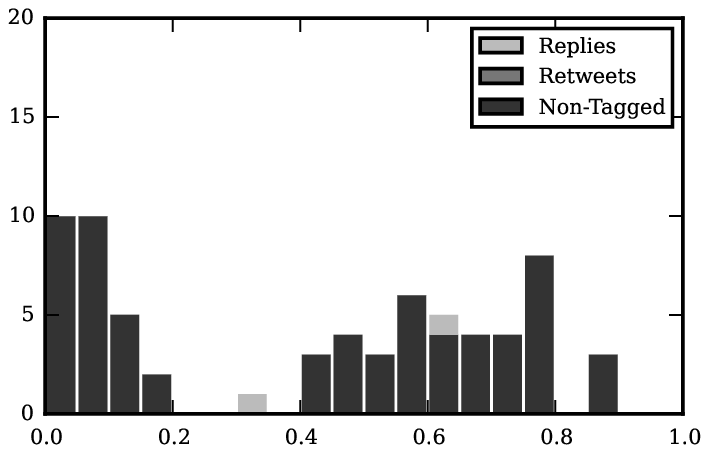}
\includegraphics[scale=0.55]{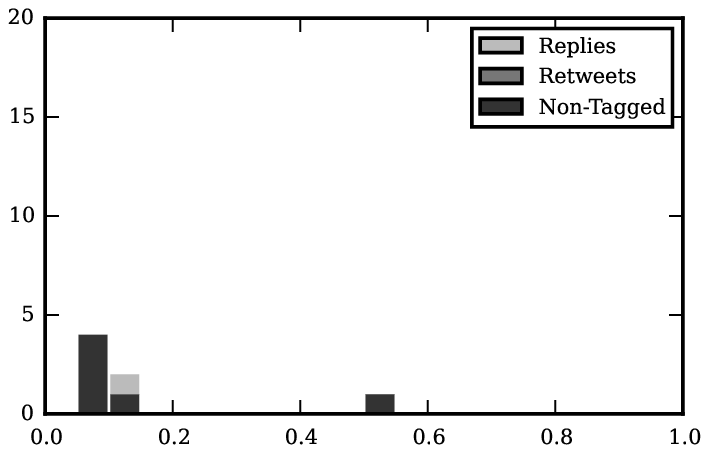}
\caption{Score histograms for sample users who present a significant amount of high scored Non-Tagged content relative to their total amount of messages, which indicates that most of their reactions are not being properly tagged by Twitter.}
\label{fig:users_reponse_histograms}
\end{figure}

These Non-Tagged high-scored messages were authored by \highUserCount{} of the \totalUsers{} users (\highUserCountPct{}\%). This suggests that users generate responses that are missed in a non-uniform way: many users behave as the system expects, using explicit reply and retweet mechanisms, but a significant number respond, at least sometimes, without using those mechanisms. 
Figure \ref{fig:users_reponse_histograms} shows histograms for example users that have most or all of their responses untagged by Twitter even though they present a high $score$.  Note that these users span a range of activity levels, meaning that they are not just newbies that don't know how to use the interface.

\begin{figure}[!tb]
\centering
\includegraphics[scale=1.2]{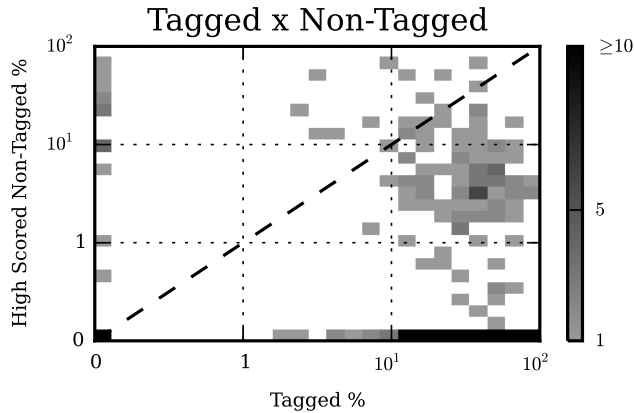}
\caption{2D histogram of the percentage of Tagged and high-scored Non-Tagged messages for all users. The scale is linear in the interval $[0,1]$ and logarithmic on the interval $(1,100]$; the dashed line represents an equal percentage of Tagged and Non-Tagged tweets. 
Many users are non-responsive (the point at the origin) or use the explicit response mechanisms consistently (points hugging the x-axis with a 0 value for high scored Non-Tagged \%).  However, a significant number never use the explicit response mechanisms (points hugging the y-axis with a 0 value for Tagged \%), use them only occasionally (points above the dashed line), or occasionally forget to use them (points below the dashed line).}
\label{fig:2d_histogram}
\end{figure}

In order to better understand the behavior distribution among all users, Figure \ref{fig:2d_histogram} shows a 2d-histogram for the points $(p_i^T, p_i^N)$, where each of these points is the percentage of the Tagged messages $p_i^T$ and the percentage of the high scored Non-Tagged messages $p_i^N$.  Each of these points is evaluated for a user $u_i$ in relation to the total number of messages the user authored.
The high scored Non-Tagged percentage $p_i^N$ is the proportion of this user behavior that were likely to be reactions while the percentage $p_i^T$ is the proportion of reactions actually captured.

The \usersScoreZero{} users that never have messages that scored higher than $\thresholdScore{}$ nor used explicit system reply mechanisms are concentrated at the origin of the histogram.
Users that lay on the $x$-axis only react through explicit reaction mechanisms the system offers, therefore have all their reaction Tagged. Similarly, users on the $y$-axis never use explicit reaction mechanisms, although they present high scored Non-Tagged content. Users above the dashed line have more high scored Non-Tagged content than Tagged content. It is possible to say that users that lay above the dashed line are more likely to produce content that can be missed by Twitter's tagging system, and they account for \usersAboveLine{} users, about $\usersAboveLinePct{}\%$ of the dataset. 

\begin{figure}[!tb]
\centering
\includegraphics[scale=1]{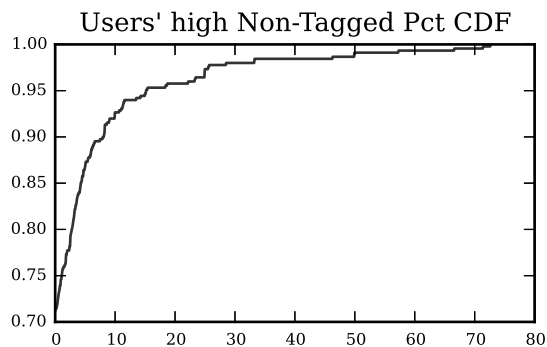}
\caption{Cumulative distribution of the users for the percentage of high scored Non-Tagged messages. \usersZeroHighScoredPct{}\% of the users have no high scored Non-Tagged messages, while \usersMoreThanTenPctPct{}\% of the users had at least 10\% of their messages high scored and Non-Tagged.}
\label{fig:cumulative_user}
\end{figure}

When considering the cumulative distribution of the users according to the percentage of high scored Non-Tagged messages $p_i^N$, shown in Figure \ref{fig:cumulative_user}, we identify more than \usersMoreThanTenPctPct{}\% of the users with at least 10\% of their messages being high scored and missed by Twitter's tagging system. 

These results indicate that methods that rely on explicit indicators of response likely miss or seriously under-represent the behavior of a sizable proportion of the Twitter population.

\section{Conclusion and Future Work}

This paper presented a novel method of capturing some of a user's non-explicit reactions to followees' content in Twitter by using text similarity scores between a user's tweets and those of their followees.  The analysis indicates that the method does generate higher scores on average for system tagged Replies and Retweets than Non-Tagged tweets, suggesting that it captures real signal about responses.  Using a conservative cutoff for predicting whether a non-tagged tweet is a response suggests that at least \highNonTaggedTweetCountPct{}\% of actual responses are not tagged by the system.  These responses are distributed across almost a quarter of the users in the dataset, with a quarter of those having more missed reaction messages than explicit system tagged ones. These are not just naive, low-activity users who do not understand Twitter and might be ignored in analysis; a number of these users are quite active, with dozens or hundreds of tweets in a 14-day window.  

Although the method has provided useful insights into the prevalence of non-explicit replies in Twitter, it is a coarse model.  It tends to under-evaluate Replies; is more sensitive to Retweet size than desirable; likely misses a number of non-explicit responses that have lower scores but are nonetheless real responses to the feed; and doesn't address responses to content outside the feed such as views by hashtag or username.  Ongoing work aims at addressing these limitations by improving the quality of the scoring function.  One natural way of improving the scoring function is to incorporate other relevant social features highlighted by past work (Table~\ref{tab:characteristics}).  We expect that better models of language, network characteristics, and attention that build on these features would give better estimates of how people react to content produced by their followees.

Another possible unfolding research topic is how to use these reaction scores to understand the reaction patterns and estimate the individual reaction level for each user.  This is important for effective models of diffusion at all levels, from understanding when adding an individual to a follower network might be most valuable, to estimating the overall reach of an individual's network, to modeling diffusion of information in the large.  Missing \highNonTaggedTweetCountPct{}\% of responses and \usersAboveLinePct{}\% users is a substantial amount of error to bear for such models, making the identification of non-explicit responses an important problem to pursue.

\section*{Acknowledgment}

The authors are grateful to FAPESP grant \#2011/50761-2, CAPES grant \#99999.009323/2014-07, NAP eScience - PRP - USP, NSF grant 1422484, Amit Sharma for his insightful comments, and RepNerv foundation for ongoing support.

\bibliographystyle{IEEEtran}
\bibliography{escience2015,escience2015handpicked}

\appendix
\clearpage
\section{Appendix}
\tweetTableFull
{Pairs of users' tweets (top in each row) and highest scoring messages in the windows (bottom in each row) for Retweets, Replies, and Non-Tagged tweets.  Tweets were randomly selected across the range of scores in each set.}
{tweetsScores}

\end{document}